# A parameterized model for Coulomb barrier height

D.K. Swami, P. Sharma and T. Nandi
*Inter University Accelerator Centre, New Delhi-110067*

## Introduction

Coulomb barrier height is a basic parameter to describing the nuclear reactions. Recently it is found that ions produce from nuclear reactions can be used to study electron loss and capture processes. Hence determining the Coulomb barrier becomes important in both nuclear and atomic physics. The different models depend on the nucleus-nucleus potential chosen or parameters used to calculate the barrier heights. To include a few Bass potential [1], Proximity potential [2], Wood Saxon potential [3], CW potential [4], Modified Wood Saxon potential [5] etc. On the other side two experimental techniques viz. fusion excitation function and quasi elastic scattering are used. The former is influenced by fusion, elastic, inelastic, transfer etc. dynamic processes, whereas the later does not at all depend on fusion processes. Hence, the second method is likely to be more accurate than the other. In this work we plan to develop a parameterized formula from experimental results obtained from quasi elastic scattering. Since quasi elastic data spread over $Z_1Z_2=64$ to $Z_1Z_2=2460$, the formula is expected to work in the range of low $Z_1Z_2$ to high $Z_1Z_2$, where $Z_1$, $Z_2$ are the atomic number of the projectile and the target, respectively.

## Calculation, results and discussions

According to the definition of Coulomb barrier ($V_B$) as given below

$$V_B = \frac{Z_1 Z_2 e^2}{R} \quad and \quad R = R_c \left( A_1^{\frac{1}{3}} + A_2^{\frac{1}{3}} \right)$$

Where $A_1$, $A_2$ stand for the mass numbers of projectile and target nuclei, respectively, $R_C$ is the radius parameter, which varies from 1.1 to 1.4 fm in different models [6].

As clear from the above equation that $V_B$ depends on $Z_1$, $Z_2$, $A_1$ and $A_2$. Hence, the experimentally obtained $V_B$ from quasi elastic measurement can be plotted against $Z_1Z_2/(A_1^{1/3}+A_2^{1/3})$ as shown in Fig. 1

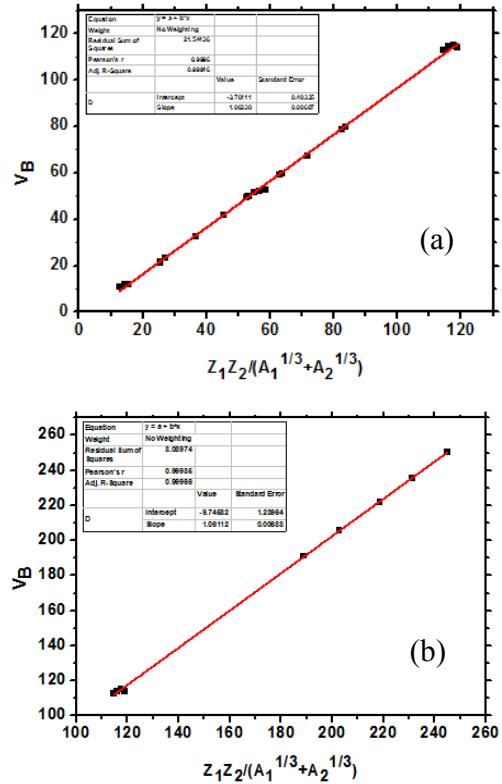

Fig. 1 Experimental $V_B$ (MeV) from quasi elastic scattering are plotted against $Z_1Z_2/(A_1^{1/3}+A_2^{1/3})$. The full set is divided into two as shown in (a) and (b), and the data are fitted with the straight line having two free parameters. The fitted functions are used as the model equation to finding $V_B$ for any systems.

It is noticed that full data set does not follow through an exact straight line where as dividing the data into two parts follows very good linearity with reduced chi square ~ 1. However, the uncertainty of the intersection parameter turns out to be large because of large errors in the measurements. Nevertheless, we fix the uncertainty within 1% on the estimated $V_B$ from the present parameterized model. These two fitting functions with two parameters are used to find the $V_B$ for any binary systems. The data so obtained follow very good agreement with the measurements from fusion excitation function as shown in Table 1. In next step, the present estimations have been compared for different binary systems with Bass and Proximity models as the representative of one dimensional barrier penetration model in Table 2. Further, the results have also been compared with the predictions from coupled channel calculation using the CCFULL code [6].

**Table 1:** $V_B$ values in MeV from fusion experiment are compared with present model along with CW and BW.

| System | Fusion | Ours | BW | CW |
|---|---|---|---|---|
| $^6$Li+$^{64}$Zn | 12.15 | 11.76±0.12 | 13.52 | 12.24 |
| $^9$Be+$^{208}$Pb | 39.49 | 37.35±0.37 | 39.49 | 37.06 |
| $^{16}$O+$^{96}$Zr | 41.36 | 41.47±0.41 | 41.88 | 40.68 |
| $^{16}$O+$^{154}$Sm | 59.71 | 59.39±0.59 | 60.29 | 58.87 |
| $^{16}$O+$^{144}$Sm | 62.61 | 60.36±0.60 | 60.95 | 59.77 |
| $^{16}$O+$^{186}$W | 67.41 | 68.43±0.68 | 69.76 | 68.12 |
| $^{48}$Ti+$^{208}$Pb | 190.1 | 190.51 | 189.90 | 187.80 |
| $^{56}$Fe+$^{208}$Pb | 223.00 | 222.26 | 221.04 | 218.53 |
| $^{86}$Kr+$^{208}$Pb | 299.2 | 293.23±0.3 | 292.61 | 287.59 |

**Table 2:** $V_B$ values in different binary systems are compared with different model predictions

| System | BW | CW | CCFULL | Proximity | Bass | Our |
|---|---|---|---|---|---|---|
| $^{16}$O+$^{16}$O | 10.52 | 9.71 | 10.26 | 10.35 | 9.74 | 8.98 |
| $^6$Li+$^{64}$Zn | 13.52 | 12.24 | 13.22 | 12.94 | 13.55 | 11.76 |
| $^{16}$O+$^{40}$Ca | 23.63 | 22.68 | 23.67 | 23.08 | 23.73 | 18.84 |
| $^{18}$O+$^{120}$Sn | 50.10 | 48.72 | 50.27 | 49.96 | 50.32 | 46.45 |
| $^{16}$O+$^{144}$Sm | 60.95 | 59.77 | 61.6 | 61.18 | 62.01 | 58.07 |
| $^{16}$O+$^{186}$W | 69.76 | 68.12 | 70.31 | 70.14 | 70.83 | 66.60 |
| $^{32}$S+$^{96}$Zr | 78.59 | 78.53 | 80.08 | 80.48 | 80.09 | 77.84 |
| $^{40}$Ca+$^{132}$Sn | 114.76 | 114.46 | 117.12 | 118.43 | 117.22 | 114.92 |
| $^{48}$Ti+$^{208}$Pb | 189.90 | 187.80 |  | 197.49 | 195.39 | 190.51 |
| $^{54}$Cr+$^{208}$Pb | 204.79 | 202.27 |  | 213.20 | 210.72 | 205.44 |
| $^{56}$Fe+$^{208}$Pb | 221.05 | 218.53 |  | 231.02 | 228.41 | 222.26 |
| $^{64}$Ni+$^{208}$Pb | 234.85 | 231.61 |  | 245.34 | 242.23 | 235.73 |
| $^{70}$Zn+$^{208}$Pb | 249.29 | 245.53 |  | 260.71 | 257.20 | 250.09 |

## Conclusion

We notice that the present model predictions are good and do not have any limitations like CCFULL predictions. This can be used to estimate reliably the Coulomb barrier height required for any experiments. This is to note that many experimental results have been used in this work the limitation space does not allow us to list them out.